\documentclass[%
 reprint,
 amsmath,amssymb,
 aps
]{revtex4-2}
\usepackage{graphicx}  
\usepackage{mwe}
\usepackage{bm,bbm,bbold,amssymb,amsmath,amsfonts,dsfont,hyperref,color}
\usepackage{mathtools}
\usepackage{physics}
\usepackage{float}
\usepackage[mathscr]{euscript}
\usepackage{array}
\usepackage{xcolor}
\usepackage{algorithm}
\floatname{algorithm}{Protocol}
\usepackage{algorithmic}

\newcommand{\average}[1]{\left\langle#1\right\rangle}
\newcommand{\qfi}[1]{\mathpzc{F}#1}
\newcommand{\re}{\mathrm{e}}
\newcommand{\ignore}[1]{}
\DeclareFontFamily{OT1}{pzc}{}
\DeclareFontShape{OT1}{pzc}{m}{it}%
              {<-> s * [1.25] pzcmi7t}{}
\DeclareMathAlphabet{\mathpzc}{OT1}{pzc}%
                                 {m}{it}

\begin{document}

\title{Private network parameter estimation with quantum sensors}
\author{Nathan Shettell}
\affiliation{Centre for Quantum Technologies, National University of Singapore, Singapore 117543, Singapore}
\author{Majid Hassani}
\affiliation{LIP6, CNRS, Sorbonne Universit\'{e}, 4 place Jussieu, F-75005 Paris, France}
\author{Damian Markham}
\affiliation{LIP6, CNRS, Sorbonne Universit\'{e}, 4 place Jussieu, F-75005 Paris, France}


\begin{abstract}
Networks of quantum sensors are a central application of burgeoning quantum networks. 
A key question for the use of such networks will be their security, particularly against malicious participants of the network. 
We introduce a protocol to securely evaluate linear functions of parameters over a network of quantum sensors, ensuring that all parties only have access to the function value, and no access to the individual parameters. This has application to secure networks of clocks and opens the door to more general applications of secure multiparty computing to networks of quantum sensors.
\end{abstract}

\maketitle

The field of quantum metrology and sensing is undergoing rapid developments in both theory and experiment \cite{giovannetti2011advances, degen2017quantum, pezze2018quantum, polino2020photonic,EscherNatPhys,Demko2015,Paris2009quantum}. By utilizing certain quantum states, one can estimate latent parameters with a super-classical level of precision \cite{pezze2018quantum, giovannetti2004, giovannetti2006, pezze2014, toth2014, huelga1997,Hassani2017}. 
Meanwhile, a quantum network \cite{chiribella2009, van2012, simon2017, kozlowski2019towards} is a collection of spatially separated quantum nodes (which may have asymmetric computational capabilities) that are interconnected through quantum channels - a means of transmitting quantum information. 
Recently, quantum networks have been shown to provide advantage for spatially distributed estimation problems \cite{komar2014, komar2016, riehle2017, eldredge2018, ge2018, zhuang2018, oh2020optimal, liu2021distributed} and multiparameter estimation problems \cite{proctor2018, qian2019,rubio2020a}.



A core problem for the way that sensors are deployed in networks is security - ensuring that the data that they collect is shared in the right way, only with authorised parties, and that it can be trusted. 
Given the incredible application of quantum technologies to cryptography \cite{pirandola2020}, it is natural to try to combine quantum cryptography with quantum sensing. Recently there have been some works in this direction \cite{komar2014,zhuang2018,xie2018,takeuchi2019,okane2020,takeuchi2019resource,SKM22,SM21}, though they are mostly restricted to settings involving two or three parties.
Here we expand this program to arbitrarily big networks of sensors, incorporating network security concerns akin to secure multiparty computation, where local inputs, or parameters, should be kept secret when cooperating to estimate a global function.

\begin{figure}[h]
    \centering
    \includegraphics[width=0.25\textwidth]{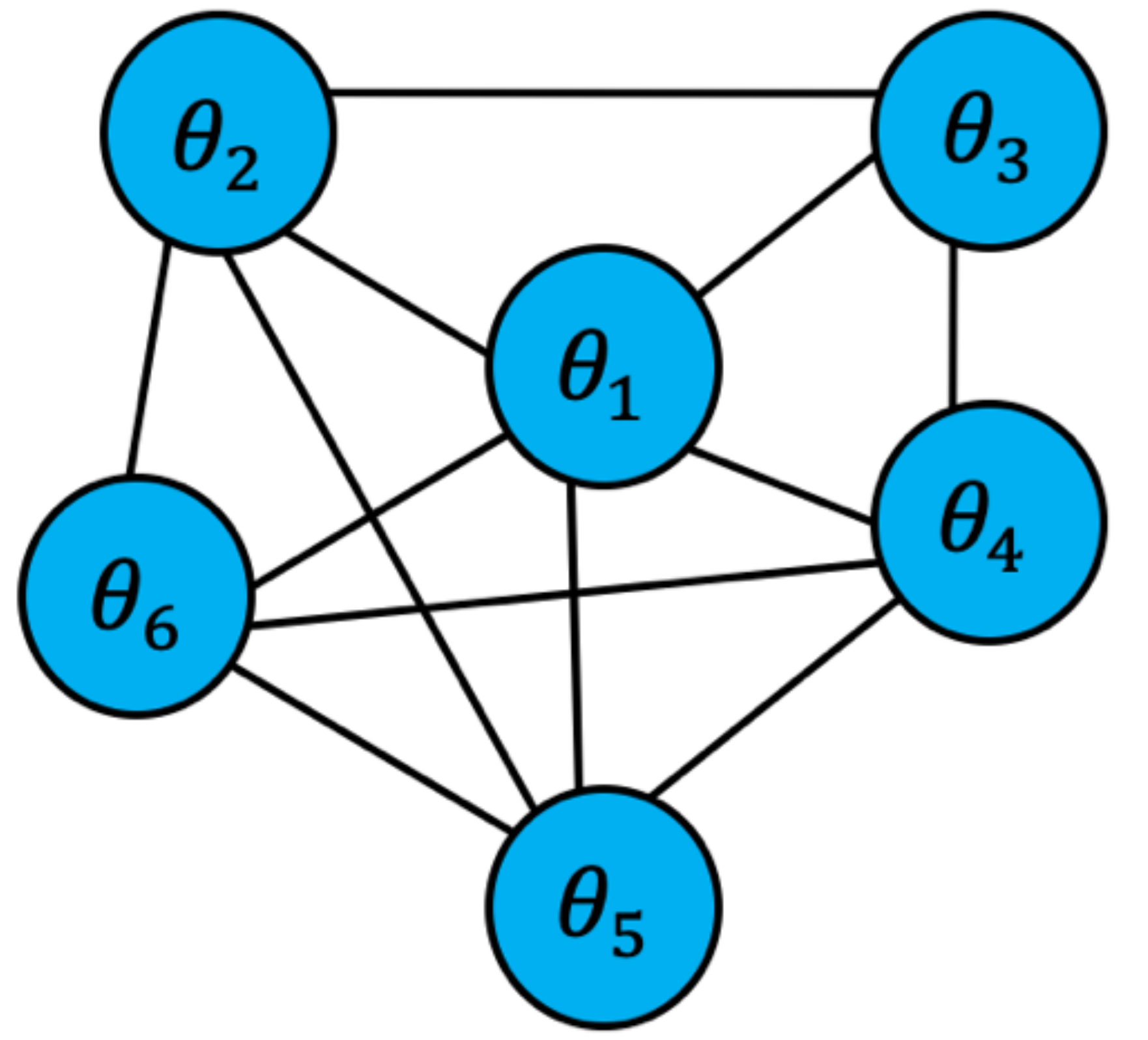}
    \caption{A quantum network with $n=6$ quantum sensors. The $j$th unknown phase is unique to the $j$th node. The quantum state is distributed throughout the quantum network via the quantum channels (represented by lines connecting the nodes). 
    We assume that the source and all of the quantum channels are untrusted, that some of the nodes may behave dishonestly and that the untrusted sources, channels and dishonest parties can work together to corrupt the network.}
\label{fig:simplenetwork} 
\end{figure}

We model a network of quantum sensors as depicted in Figure~\ref{fig:simplenetwork}, following \cite{proctor2018,rubio2020a}. The network is made of $n$ nodes representing quantum sensors, where each node encodes a parameter $\theta_i$ to all qubits at that node.
In principle these may be single parameters or vectors of parameters (encoding field strength and direction for example). For simplicity we restrict ourselves to single parameters, though the ideas and techniques present can be naturally extended.
The sensors are imagined to be connected via quantum and classical channels so they can share entangled states across the network. One may also allow in this way to perform global measurements, however in our case local measurements are sufficient.
Though one may consider different network setups (for example see \cite{sekatski2020optimal}), this setting is rich, suited to optical phase estimation \cite{polino2020photonic,liu2021distributed} mapping varying fields \cite{altenburg2017estimation}, networks of clocks \cite{komar2014} and more \cite{proctor2018,rubio2020a}, and with many applications where sharing entangled states allows for estimation of global functions more efficiently than locally estimating the parameters and communicating classically \cite{komar2014, komar2016, riehle2017, eldredge2018, ge2018, zhuang2018, proctor2018, qian2019, rubio2020a, guo2020, zhang2021}.
It is also adapted to our analysis, and we expect our techniques and approaches can be applied to other network settings straightforwardly.

We consider different security and trust assumptions over this network.
We allow for untrusted channels, sources and dishonest behaviour form nodes. Furthermore all these malicious parties are allowed to work together to coordinate their attacks on the network.
In this setting we wish to estimate function $f(\Theta)$ over the local parameters $\Theta = \{ \theta_1, \ldots, \theta_n \}$ and we will address two main security questions. First we want to be sure that our estimation of $f(\Theta)$ is good, which we call \textit{integrity}, which we quantify following \cite{SKM22,SM21}. Second, we introduce the notion of \textit{privacy} for networks of sensors, wherein we allow for the estimation only of $f(\Theta)$, but all other information about the local parameters $\Theta$ are kept secret from the rest of the network. This is analogous to the input privacy constraints in secure multiparty computing (SMPC), and is motivated by the need of protecting local information, as is commonly applied in classical networks. 
Furthermore, we will introduce an approximate version of network privacy, which we need for security statements.

We present a protocol which allows us to estimate optimally global functions of the form 
$f(\Theta) = M \sum_i k_i \theta_i$, where $k_i \in \mathbb{Z}$, such that any parties can only have access to $f(\Theta)_H = M \sum_{i\in H} i k_i \theta_i$, where $H$ is the set of honest nodes, and no other information of the local parameters is available to any party. 
We will first present the protocol first for the case where $k_i = 1/nM$, corresponding to estimating the the average value of the $n$ phases, i.e. $\bar{\theta}=\frac{\theta_1 + \ldots + \theta_n}{n}$. The generalisation to the above is straightforward. Estimation of the average phase is the simplest estimation of a linear functional \cite{rubio2020a} and has application, for example, to the synchronisation of networks of clocks such as for in satellites where the security we provide can be interpreted as preserving sovereignty of the satellite whilst allowing for clock synchronisation\cite{komar2014, komar2016}. 

\begin{figure}[h]
    \centering 
    \includegraphics[width=0.47\textwidth]{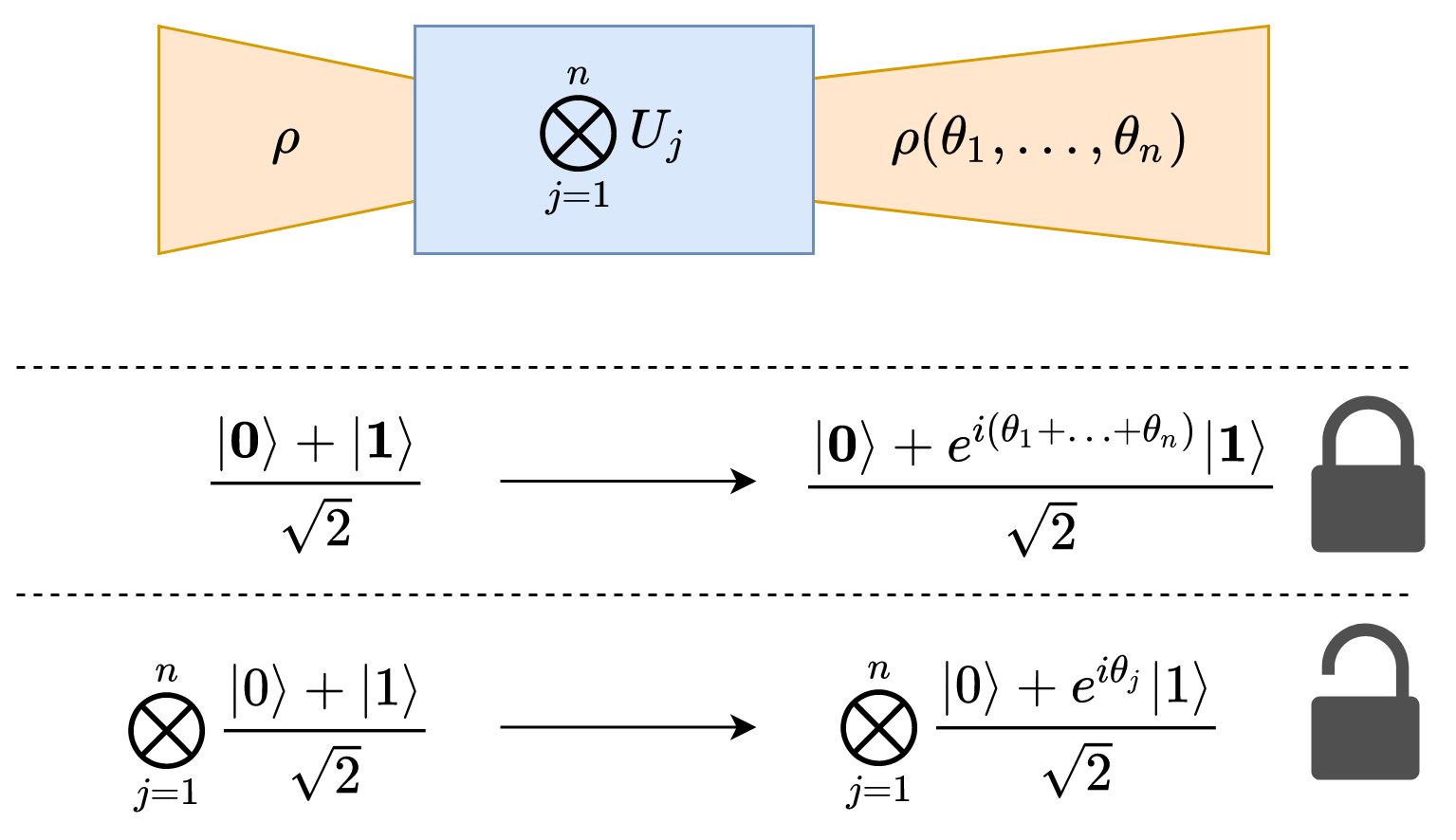}
    \caption{In a quantum sensing network, a parameter local to each individual node, $\theta_j$ is encoded into the shared quantum state $\rho$. The accessible information in the encoded quantum state $\rho(\theta_1,\ldots, \theta_n)$ is highly dependent on $\rho$. For example, one cannot estimate the value of any individual phase by using a GHZ state; this is not true however for the separable state. This is the basis of privacy.}
\label{fig:anon} 
\end{figure}

The broad idea is the following. Sharing a GHZ state allows for optimal estimation of the average of phases (\cite{komar2014,komar2016,rubio2020a}). Crucially, as we describe below, when phases are encoded into it this way, it \textit{also} has the desired security properties, related to its use for anonymous communication \cite{christandl2005quantum,brassard2007anonymous,unnikrishnan2019anonymity} - essentially all local information is erased and only the global average value remains in the state.
However if a different state is used, for example a product of Pauli X eigenstates, the local phase information is stored perfectly (see Fig.~\ref{fig:anon}). 
In order to stop malicious parties from accessing the honest parties individual $\theta_i$ in this way, our strategy is to certify the resource state that will be used for sensing is close to a GHZ state.
For this we apply tests developed for the presence of dishonest parties \cite{unnikrishnan2020verification}. Then, from continuity of the quantum Fisher information \cite{augusiak2016asymptotic,majid19} we will see that being close to the GHZ state implies that we both have the information we want, and no more.

Our first tool, then, is to use the verification of graph states in \cite{unnikrishnan2020verification}, for the case of the GHZ state, which is locally equivalent to the graph state for the star or the complete graph (see also \cite{pappa2012multipartite} for a similar verification protocol for the GHZ state).
As with many similar verification protocols it is based on checking stabiliser conditions. 
In particular, for the  $n$-party GHZ state we define the following stabiliser generators
\begin{equation}
    \begin{split}
        K_1 = -YYX &\ldots X X X , \\
        K_2 = -XYY &\ldots X X X, \\
        &\;\; \vdots \\
        K_{n-2} = -XXX &\ldots X Y Y, \\
        K_{n-1} = -YXX &\ldots X X Y, \\
        K_n = XXX &\ldots XXX,    \end{split}
    \label{eq:GHZgenerators}
\end{equation}
which stabilise the state in so far as $K_i |GHZ_n\rangle=|GHZ_n\rangle$. 
In Protocol \ref{protocol: Verification} we present the version of the protocol with one party identified as the Verifier, which is assumed to be honest. The protocol can be adapted to allow for dishonest Verifiers following \cite{unnikrishnan2020verification} by using a trusted Common Random Source. This is presented in the appendix, with the slightly adapted security statements.

\begin{figure}
\begin{algorithm}[H]
\caption{\textsc{Verification} \cite{unnikrishnan2020verification}} 
\begin{flushleft}
\textit{Input}: The parties choose values of $N_{test}, N_{total}$. Let $\mathcal{S}$ be the set of test measurements, and $\mathsf{J} = \abs{\mathcal{S}}$.  \\
\textit{Goal}: Distribution of a $GHZ$ state.\\

\end{flushleft}
\begin{algorithmic}[1]
\STATE The Verifier requests $N_\text{total}=2 n N_\text{test}$ copies of $\ket{\psi}$, where $N_\text{test}$ is the number of tests done per generator (see Fig.~\ref{protocoloutline}). \\ \

\STATE The total state $\ket{\psi}^{\otimes N_\text{total}}$ is distributed throughout the quantum network, where $j$th qubit of each copy is sent to the $j$th node.

\STATE For each $1 \leq j \leq n$ Eq.~\eqref{eq:GHZgenerators}: \vspace{-5pt}

\begin{enumerate}

        \item The Verifier randomly selects $N_\text{test}$ copies of $\ket{\psi}$ to be measured with respect to $K_j$, after which it is discarded (i.e. a measured copy cannot be re-used for a later test).
        
        \item All $n$ nodes send the measurement results to the Verifier.
        
        \item The $j$th failure rate (i.e. the number of tests which resulted in a $-1$ eigenvalue of $K_j$ divided by $N_\text{test}$) is recorded as $f_j$.
        
    \end{enumerate}

\STATE  The Verifier randomly selects a copy from the remaining $N_\text{total}-n N_\text{test}=N_\text{total}/2$ non-tested copies. This is known as the target copy, denoted by $\tilde{\rho}$. All other copies are discarded. \\ \

\STATE The average failure rate $f=\frac{1}{n}\sum_j  f_j$ is computed. If $f \leq \frac{1}{2n^2}$, then $\tilde{\rho}$ is output for use, otherwise it is discarded.
    
\end{algorithmic}
\label{protocol: Verification}
\end{algorithm}
\end{figure}

Protocol \ref{protocol: Verification} outputs a state which is certified to be close to the GHZ state, over the honest parties ($H$).  Note that this is clearly the best one could hope for as dishonest parties can do whatever they want to their part of the state.
In \cite{unnikrishnan2020verification}, it is shown that if one chooses positive constants $m$ and $c$ such that $\frac{3}{2m} < c < \frac{(n-1)^2}{4}$ and sets $N_\text{test}=\lceil m n^4 \ln n \rceil$, then 
\begin{equation}
    \mathbb{P} \Big( F(\tilde{\rho}^{(H)},\dyad{\psi}^{(H)}) \geq  1-\frac{2 \sqrt{c}}{n}-2nf \Big) \geq 1-n^{1-\frac{2mc}{3}},
    \label{eq:protocolfidelity}
\end{equation}
where $\mathbb{P}(A)$ is the probability of (A) being true, $F$ is the fidelity and the superscript $(H)$ denotes the reduced state over all of the honest parties; notice that there is an inherent trade-off between the probability and fidelity bounds. In this work we use the squared version of fidelity, $    F(\rho_1,\rho_2) = \Tr ( \sqrt{\rho_1} \rho_2 \sqrt{\rho_1})^2$, to remain consistent with previous work \cite{SKM22}.

We now move to how the output verified GHZ state is used for the sensing part, and how it ensures privacy against malicious participants.
The encoding of the average of the phases onto the GHZ goes as follows.
The $j$th phase, $\theta_j$, is encoded at the $j$th node via unitary evolution $U_j=\dyad{0}+{\re}^{i\theta_j} \dyad{1}$. 
After distributing a GHZ state, $\ket{\psi}$, throughout the quantum network, the overall unitary evolution caused by $\bigotimes_j U_j$ is
\begin{align}
    |GHZ_n\rangle &= \frac{\ket{\bm{0}}+\ket{\bm{1}}}{\sqrt{2}} \xrightarrow[]{\bigotimes_j U_j} \frac{\ket{\bm{0}}+e^{i n \bar{\theta}} \ket{\bm{1}}}{\sqrt{2}},
\end{align}
where $\ket{\bm{x}}=\ket{x}^{\otimes n}$. After the unitary encoding, the average phase $\bar{\theta}$ can be estimated with a super-classical level of precision known as the Heisenberg limit \cite{giovannetti2004, giovannetti2006}.

Turning to privacy, one can see from the form of the encoded state that it only contains information of the average of the phases, and all other finer details are lost.
Indeed for any given $\theta_i$ one can hide its value totally by choosing different values of $\theta_{j\neq i}$. 
Motivated by this we say that an encoding is \textit{private} if for any $j$ there exists a set of values for $\theta_k \rightarrow \phi_k$, $k \neq j$, such that the encoded state is independent of $\theta_j$ when $\theta_k$ take on the prescribed values. For example, the encoded GHZ state is perfectly private as one can set $\phi_k = - \frac{\theta_j}{n-1}$ and obtain a quantum state independent of $\theta_j$,
\begin{align}
    \label{eq:GHZanonymous}
    \frac{\ket{\bm{0}}+e^{i n \bar{\theta}} \ket{\bm{1}}}{\sqrt{2}} & \xrightarrow[]{\theta_k \rightarrow -\frac{\theta_j}{n-1} \forall k \neq j}  \frac{\ket{\bm{0}}+ \ket{\bm{1}}}{\sqrt{2}}.
\end{align}
In order to quantify security in the context of sensing, we use the quantum fisher information,
$\qfi$, as it is a natural metric which quantifies the ultimate amount of information a quantum state contains about an unknown parameter(s) \cite{barndorff2000}. An overview of the QFI is given in the appendices. 
A general quantum state $\rho$, which has been encoded with multiple parameters $\{ \theta_1, \ldots, \theta_n \}$, is perfectly private if $ \forall j$ there exists a choice of $\phi_k$ $\forall k \neq j$ such that
\begin{equation}
    \label{eq:anonymity}
    \qfi[\rho \big|_{\theta_k \rightarrow \phi_k \forall k \neq j}] = 0.
\end{equation}
We say an $n$ qubit state $\rho$, which has been encoded with multiple parameters $\{ \theta_1, \ldots, \theta_n \}$ is $\varepsilon$\textit{-private} if $\forall j$ there exists a choice of $\phi_k$ $\forall k \neq j$ such that
\begin{equation}
    \label{eq:approxanonymity}
    \qfi[\rho \big|_{\theta_k \rightarrow \phi_k \forall k \neq j}] \leq \varepsilon n^2.
\end{equation}
The scaling of $n^2$ is meant to reflect the maximum achievable value the QFI can take and consequently restricting $\varepsilon \in [0,1]$. In settings which this is not the case (e.g. problems other than phase estimation or using qudits instead of qubits), the scaling factor can be changed.



We now combine the verification with the sensing in Protocol \ref{protocol: secure sensing}.

\begin{figure}[h]
\begin{algorithm}[H]
\caption{\textsc{Secure network sensing}} 
\begin{flushleft}
\textit{Input}: The parties choose value of $\nu$, and have local parameters $\Theta = \{\theta_1, ... \theta_n\}$.  \\
\textit{Goal}: Estimate function $f(\Theta)$, with $\epsilon$-integrity and $\epsilon$-privacy.\\

\end{flushleft}
\begin{algorithmic}[1]
\STATE The nodes run \textsc{VERIFICATION} to share a GHZ state \\ \
\STATE   Each node $j$ encodes their local parameter $\theta_j$ by applying $U_j = e^{i \theta_j}$ to all of their qubits\\ \
\STATE   Each node measures their qubits in X and announces the parity of their results publicly\\ \
\STATE   Steps 1-3 are repeated $\nu$ times and the public data used to estimate $f(\Theta)$\\ \

\end{algorithmic}
\label{protocol: secure sensing}
\end{algorithm}
\end{figure}

Following \cite{SKM22,SM21} the closeness of the resource state from the verification protocol guarantees that the estimation can be trusted. This is captured by the following theorem, which follows directly from (\ref{eq:protocolfidelity}) and \cite{SKM22,SM21}.

\bigskip

\textit{Theorem 1: Integrity in the honest case.} \\
Let $\hat{f(\Theta)}$ denote the estimate of function $f(\Theta)$ for the ideal resource state and let $\hat{f(\Theta)}^\prime$ denote the actual estimation coming from Protocol \ref{protocol: secure sensing}. 
If all nodes behave honestly, then with probability at least $1-n^{1-\frac{2mc}{3}}$, the estimate  bias of the function is bounded via
\begin{equation}
\big| \mathbb{E}(\hat{f(\Theta)}^\prime) - \mathbb{E}(\hat{f(\Theta)}) \big|  \leq \frac{2o\varepsilon}{|\partial \expval{O}_{\rho_\theta}|}
\label{eqn:AccBias}
\end{equation}
and the difference in precision is bounded via
\begin{equation}
\big| \Delta^2\hat{f(\Theta)}^\prime-\Delta^2\hat{f(\Theta)} \big| \leq  \frac{4 o^2 (2\varepsilon \nu^{-1}+\varepsilon^2 )}{|\partial \expval{O}_{\rho_\theta}|^2},
\end{equation}
where $\epsilon \leq \frac{2\sqrt{c}+1}{n}$ and $o$ is the maximum magnitude of the eigenvalues of $O$, which is the observable measured to perform the estimation of the function - for estimating $\bar{\theta}$ with a GHZ state $O=X^\otimes{n}$, hence $o=1$.

\bigskip

Theorem 1 is true even if the channels and source are untrusted, but we note that it makes sense only if all nodes are honest - since a dishonest node can always change the global estimation of the function if they wish.
If one trusts all the nodes, one could also ensure that no eavesdropper could access the estimation of the function following \cite{SKM22,SM21}. 
However, we assume that we cannot know who the dishonest parties are (otherwise we may simply exclude them from the network), making privacy more subtle.

To address privacy, we next use the the continuity of the QFI \cite{augusiak2016asymptotic}, which for this instance can be written as
\begin{equation}
    \label{eq:continuity}
   \big| \qfi[\rho_1]- \qfi[\rho_2] \big| \leq 24 \sqrt{1-F(\rho_1,\rho_2)},
\end{equation}
as shown in the Appendix. Note that a tighter inequality can be derived if one knows the rank of $\tilde{\rho}^{(H)}_\text{enc}$ \cite{majid19}. By combining Eq.~\eqref{eq:protocolfidelity}, Eq.~\eqref{eq:anonymity} and  Eq.~\eqref{eq:continuity}, one can determine a bound on the privacy. 
As for Eq.~\eqref{eq:protocolfidelity}, we are restricted to the reduced state of the honest parties. This is because we cannot certify whether or not a malicious node has encoded their respective unknown parameter. Therefore, the only thing which one can assure is a notion of privacy with respect to the unknown parameters of the honest nodes, we denote this set as $\boldsymbol{\theta}^{(H)}$, from which we arrive at the following theorem.\\

\bigskip

\textit{Theorem 2: Privacy.} \\
For all $j$ such that $\theta_j \in \boldsymbol{\theta}^{(H)}$ we have that there exists a choice of $\phi_k$ $\forall k \neq j$ (and $\theta_k \in \boldsymbol{\theta}^{(H)}$) such that
\begin{equation}
    \mathbb{P} \Big( \qfi[\tilde{\rho}^{(H)}_\text{enc} \big|_{\theta_k \rightarrow \phi_k \forall k \neq j}] \leq  24 \sqrt{\frac{2 \sqrt{c}}{n}-2nf}  \Big) \geq 1-n^{1-\frac{2mc}{3}}.
    \label{eq:protocolanonymity}
\end{equation}
Hence, Protocol \ref{protocol: secure sensing} guarantees $\varepsilon=\frac{24}{\tilde{n}}\sqrt{\frac{2 \sqrt{c}}{n}-2nf}$-privacy with probability of at least $1-n^{1-\frac{2mc}{3}}$, where $\tilde{n}$ is the number of honest parties. 

\bigskip

In the above we have treated the estimation of the average of phases. To extend to more general functions we can easily do so simply by grouping qubits, and applying Pauli $X$ operations to the resource state.
In particular to encode the function $f(\bar\Theta) = M \sum_i k_i \theta_i$, where $k_i \in \mathbb{Z}$, we start with a GHZ state of $\sum_i |k_i|$ qubits, grouped into sets of $|k_i|$. When $k_i$ is negative we apply a Pauli X on those $k_i$ qubits. The qubits $|k_i|$ of this resource state are sent to node $i$, so that all $k_i$ of them undergo the same unitary encoding $e^{i\theta_i Z}$ during the encoding phase. Since the resource state is locally equivalent to the GHZ state, the verification protocol runs exactly the same, up to the local $X$ Paulis, and the security follows similarly.


\bigskip 

In this work we have introduced the notion of privacy for networks of quantum sensors as denoting the capacity to estimate a global function of the parameters, without revealing value of local parameters themselves. We define an approximate version of this quantified by the quantum Fisher information. We then apply existing verification techniques to give a protocol for private parameter function estimation, with two security properties. Firstly, if all nodes behave honestly the estimation is good (integrity), secondly, dishonest nodes, even with collaboration of the untrusted channels and sources cannot get information of the local parameters more than that in the function to be estimated (privacy).

Our protocol is very practical and could be implemented easily for example in optical set ups. Indeed similar verification protocols have been carried out \cite{bell2014experimental,mccutcheon2016experimental} in systems set ups which are similar to those used for optical sensing networks \cite{liu2021distributed}. 

Our approach is also easily adaptable to other situations. 
Different verification protocols and techniques can be plugged into Protocol 2. This can be useful for example if want to use resource states other than graph states, or to change the interaction used, for example if quantum interaction is not possible with the sensors, one may use protocols based on classical complexity assumptions \cite{metger2021self}.
The notion of privacy we present works for different approaches to metrology such as the Baysian approach also, though quantification would have to be reassessed. 
It is natural to try to extend this approach to more general linear functionals \cite{eldredge2018, qian2019, zhang2021} or multiparameter problems \cite{ge2018, proctor2018, rubio2020a}. In this respect it is interesting to note that sensing advantage coincides with security advantage here. We leave these questions to future work.
As the quantum internet becomes a reality, allowing for networks of sensors to be deployed, this work opens up many opportunities for exploring and exploiting privacy.





\textit{Acknowledgements.---} We acknowledge funding from the ANR through the ANR-17-CE24-0035 VanQuTe project and EU/ANR via QuantERA/ShoQC.

\bibliographystyle{unsrtnat}
\bibliography{main}
\vfill \pagebreak \onecolumngrid
\appendix
\section{Quantum Fisher Information}
\noindent Estimation of unknown parameter of quantum dynamics is the main purpose of quantum metrology and sensing \cite{CavesBraunstein,Paris2009quantum}. In order to do that, one prepares a proper initial quantum state as a probe, $\varrho _{0}$, (preparation stage) and interacts it with quantum dynamics (sampling stage). Finally the probe, $\varrho _{\theta}$, is measured by a set of positive operators (measurement stage). The quantum estimation theory provides useful tools to extract the value of unknown parameter from the outcomes of measurement. 
For any locally unbiased estimator $\hat{\theta}$ for $\theta$, the ultimate precision limit can be addressed by the Cram\'{e}r-Rao bound 
\begin{equation}\label{CRB}
\text{Var}(\hat{\theta})\geqslant\frac{1}{\qfi[\varrho _{\theta}]},
\end{equation}
where $ \text{Var}(\hat{\theta}) $ is the variance of the estimator and $ \qfi[\varrho _{\theta}] $ denotes the quantum Fisher information (QFI) which quantifies the ultimate extractable amount of information about the unknown parameter, $\theta$, from the probe state, $\varrho _{\theta}$, over all possible measurements. The QFI is given by
\begin{equation}\label{qfi}
\qfi[\varrho _{\theta}]=\Tr[\varrho _{\theta} L_{\theta}^{2}]\overset{(\ref{slddef})}{=}\Tr[\partial _{\theta}\varrho _{\theta} L_{\theta}],
\end{equation}
where $ L_{\theta} $ denotes the symmetric logarithmic derivative (SLD) which is defined as 
\begin{equation}
    \label{slddef}
    \partial _{\theta}\varrho_{\theta}=\frac{L_{\theta}\varrho_{\theta} +\varrho_{\theta} L_{\theta}}{2},
\end{equation}
and $\partial _{\theta}$ shorthanded for $\frac{\partial}{\partial\theta _{\mu}}$. The optimal measurement which saturates the Cram\'{e}r-Rao bound can be constructed by the eigenvectors of the SLD. In what follows, we calculate the SLD and the QFI for two general classes of states: pure states and full rank density matrices. 

\noindent Utilizing the fact that for a pure state $\varrho _{\theta}^{2}=\varrho _{\theta}=\dyad{\psi _{\theta}}$ and $(\partial _{\theta}\varrho _{\theta})\varrho _{\theta}+\varrho _{\theta}(\partial _{\theta}\varrho _{\theta})=\partial _{\theta}\varrho _{\theta}$, one obtains
\begin{equation}
    \label{sldpure}
    L_{\theta}=2(\ket{\partial _{\theta}\psi _{\theta}}\bra{\psi _{\theta}}+\ket{\psi _{\theta}}\bra{\partial _{\theta}\psi _{\theta}}).
\end{equation}
Whence the QFI of each pure state is equal to
\begin{equation}
    \label{qfipure}
    \qfi[\dyad{\psi _{\theta}}]=4\left(\average{\partial _{\theta}\psi _{\theta}|\partial _{\theta}\psi _{\theta}}-\vert\average{\psi _{\theta}|\partial _{\theta}\psi _{\theta}}\vert ^{2}\right).
\end{equation}
By exploiting the spectral decomposition for a full rank density matrix $\varrho _{\theta}=\sum_{k}\lambda _{k}\dyad{k}$, where $0<\lambda_{k}\leqslant 1 $, Eq.~(\ref{slddef}) gives
\begin{equation}
    \label{sldfullrank}
    L_{\theta}=2\sum _{k,l}\frac{\average{k|\partial _{\theta}\varrho _{\theta}|l}}{\lambda _{k}+\lambda _{l}}\ket{k}\bra{l}.
\end{equation}
From Eq.~(\ref{qfi}), the QFI for any full rank density matrix has a closed form expression 
\begin{equation}
    \label{qfifullrank}
    \qfi[\varrho _{\theta}]=2\sum _{k,l}\frac{\vert\average{k|\partial _{\theta}\varrho _{\theta}|l}\vert ^{2}}{\lambda _{k}+\lambda _{l}}.
\end{equation}
\section{continuity proof}\label{continuityproof}
\noindent The continuity of QFI has been broadly studied. In Ref.~\cite{Safranek} the discontinuity of the QFI and the Bures metric have been explored in the sense that the unknown parameterization of the quantum state changes infinitesimally. In Ref.~\cite{majid19} the continuity property of the QFI has been studied in a general case without any assumption on the form of quantum dynamics and initial states. The continuity of the QFI for special case of unitary evolutions has been demonstrated in \cite{augusiak2016asymptotic} which is more compatible with our current purpose. For any mixed state $\varrho _{0}$ and any pure state $\sigma _{0}$ evolved under the unitary operation in which the unknown parameter is encoded multiplicatively, namely $U_{\theta}=\re ^{-i\theta H}$, the continuity relation of the QFI is expressed as~\cite{augusiak2016asymptotic}
\begin{equation}
    \label{continuity-gen}
    \vert\qfi[\varrho_{\theta}]-\qfi[\sigma_{\theta}]\vert\leqslant 24\Vert H\Vert_{\infty}^{2}\sqrt{1-F(\varrho _{\theta},\sigma _{\theta})},
\end{equation}
where $\Vert A\Vert _{\infty}$ is the largest eigenvalue of the matrix A and $F(\varrho _{\theta},\sigma _{\theta})$ denotes the fidelity. In our case of interest the unitary evolution is equal to
\begin{equation}
    \label{genunitary}
    U=\text{exp}(\sum _{j=1}^{n} H _{j}),
\end{equation}
in which $H_{j}=\mathbb{1}\otimes\mathbb{1}\cdots\otimes(-i\theta _{j}H)\otimes\cdots\otimes\mathbb{1}$, $\sigma _{0}=\dyad{\psi}$, and $\varrho _{0}$ is an arbitrary initial state. Applying the definition of privacy (by replacing $\theta _{j\neq k}=\frac{-\theta _{k}}{n-1}$) to Eq.~\ref{genunitary} yields
\begin{equation}
    \label{anoym-unitary}
    U=\text{exp}(-i\theta _{k} H'),
\end{equation}
where
\begin{equation}
    \label{Hprime}
    H'=\frac{-H}{n-1}\otimes\mathbb{1}\cdots\otimes\mathbb{1}+\cdots+\underbrace{\mathbb{1}\cdots\otimes H\otimes\cdots\otimes\mathbb{1}}_{k^{\text{th}}~\text{term}}+\cdots+\mathbb{1}\otimes\mathbb{1}\cdots\otimes\frac{-H}{n-1}.
\end{equation}
In order to calculate the standard norm of $H'$, we apply the triangle inequality
\begin{equation}
    \label{normH}
    \Vert H'\Vert _{\infty}\leqslant 2\Vert H\Vert_{\infty}=1.
\end{equation}
In the last equality we used the fact that $\Vert H\Vert _{\infty}=\Vert\sigma _{z}\Vert _{\infty}/2=1/2$. Substituting  Eq.~\ref{normH} in Eq.~\ref{continuity-gen} yields
\begin{equation}
    \label{anoym-continuity2}
     \vert\qfi[\varrho_{\theta}]\vert\leqslant 24\sqrt{1-F(\varrho _{\theta},\sigma _{\theta})}.
\end{equation}
Note that applying the definition of privacy vanishes the QFI of the GHZ state.


\section{Symmetrised verification protocol}

\begin{figure}[h]
    \centering
    \includegraphics[width=0.38\textwidth]{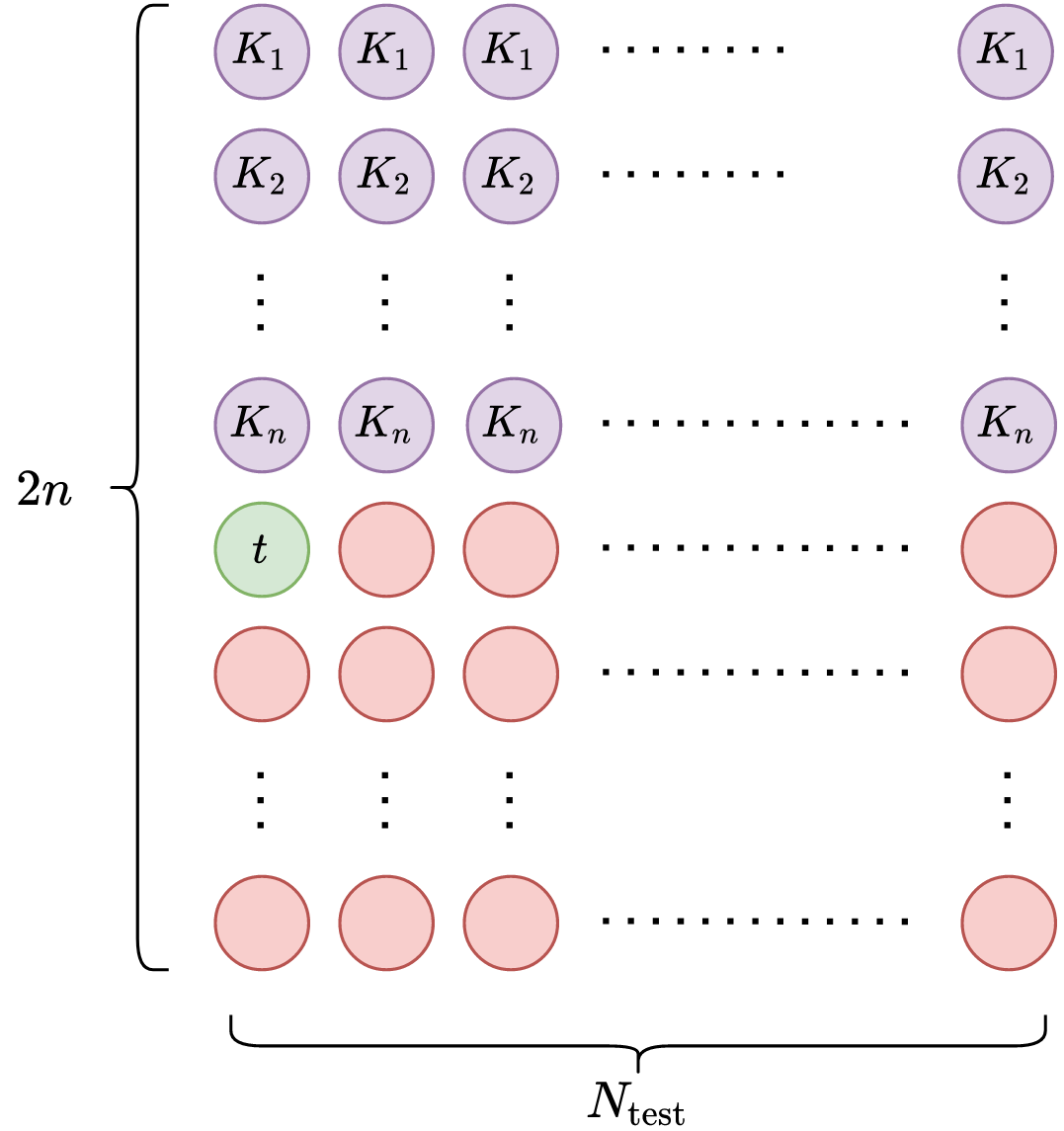}
    \caption{To certify that the distributed quantum state is in fact the GHZ state, the external source generates $N_\text{total}=2n N_\text{test}$ copies of the GHZ state. For each generator $K_j$, Eq.~\eqref{eq:GHZgenerators}, $N_\text{test}$ copies are used for the respective stabilizer measurement in order to detect and thwart any malicious interference. From the remaining $N_\text{total}/2$ untested copies, one is randomly selected as the target copy, which is to be used for the quantum sensing task. Note that which copies serve which purpose are randomly selected, the uniform ordering illustrated is intended for clarity.}
\label{protocoloutline} 
\end{figure}

In the main text, the verification protocol assumes an honest Verifier runs the protocol. In some cases one may wish to also allow for a dishonest verifier. Following \cite{unnikrishnan2020verification} this is achieved by increasing the number of tested copies from $N_\text{test}$ to $\lambda N_\text{test}$ and the use of a trusted Common Random Source (CRS) which replaces the random choices of the Verifier, and for the comparison of the results chooses a Verifier at random so that there is a non-negligible proportion of times when the Verifier is honest, hence the protocol works.

The symmetrised version of the protocol runs as follows \cite{unnikrishnan2020verification}

\begin{enumerate}
    \item The Verifier generates $N_\text{total}= (\lambda + 1) \lambda  n N_\text{test}$ copies of $\ket{\psi}$, where $N_\text{test}$ is the number of tests done per generator (see Fig.~\ref{protocoloutline}).
    
    \item The total state $\ket{\psi}^{\otimes N_\text{total}}$ is distributed throughout the quantum network, where $j$th qubit of each copy is sent to the $j$th node.
    
    \item For each $1 \leq j \leq n$ Eq.~\eqref{eq:GHZgenerators}: \vspace{-5pt}
    
    \begin{enumerate}
        \item The CRS  randomly selects $\lambda N_\text{test}$ copies of $\ket{\psi}$. These are randomly split into $\lambda$ sets of $N_\text{test}$, again at random by the CRS. For each $N_\text{test}$ copies, the CRS chooses a random node to be the Verifier. Each copy should be measured with respect to $K_j$, after which it is discarded (i.e. a measured copy cannot be re-used for a later test).
        
        \item All $n$ nodes send the measurement results to the Verifier for the corresponding copy.
        
        \item The $j$th failure rate (i.e. the number of tests which resulted in a $-1$ eigenvalue of $K_j$ divided by $N_\text{test}$) is recorded as $f_j$.
    \end{enumerate}
    
    \item The Verifier randomly selects a copy from the remaining $N_\text{total}-\lambda n N_\text{test}$ non-tested copies. This is known as the target copy, denoted by $\tilde{\rho}$. All other copies are discarded.
    
    \item The average failure rate $f=\frac{1}{\lambda n}\sum_j  f_j$ is computed. If $f \leq \frac{1}{2\lambda n^2}$, then $\tilde{\rho}$ is used for the quantum sensing task, otherwise it is discarded.

\end{enumerate}

Then, setting $N_\text{test}=\lceil m n^4 \ln n \rceil$, and choosing positive constants $m$ and $c$ such that the probabilities remain positive, we have \cite{unnikrishnan2020verification}
\begin{equation}
    \mathbb{P} \left( F(\tilde{\rho}^{(h)},\dyad{\psi}^{(h)}) \geq  1- \left(\frac{1}{\lambda} -\frac{1}{\lambda^2}\right) - \left(1+\frac{1}{\lambda}\right)\left( \frac{\sqrt{c}}{n} + \lambda n f   \right) \right) 
    \geq 1- \sum_{x=0}^\lambda \left( 1 - \frac{|H|}{n}\right)^x\left(\frac{|H|}{n}n^{\frac{-2cm}{3}}\right)^{\lambda-x}.
    \label{eq:symmprotocolfidelity}
\end{equation}

One can apply this in exactly the analogous way to the asymmetric protocol to sensing to get analogous statements for the integrity and privacy.

\end{document}